\documentclass{elsart}

\usepackage{graphicx}

\usepackage{amssymb}

\begin{document}
\begin{flushright}
\hfill SMU-HEP-05-10
\end{flushright}

\begin{frontmatter}



\title{Simple acoustical technique for automated measurement of drift tube anode wire
tension}


\author{Michael Hosack} and \author{Thomas Coan\corauthref{cor}}
\corauth[cor]{Corresponding author. Tel. (+1) 214 768 2497, fax (+1) 214 768 4095,
email: coan@mail.physics.smu.edu}
\address{Physics Department, Southern Methodist University, Dallas, TX 75275, USA}

\begin{abstract}
We describe a simple and inexpensive acoustical technique that permits
rapid, accurate and in-situ measurement of drift tube anode wire
tensions even if the anode wire is electrically discontinuous.

\end{abstract}

\end{frontmatter}

\section{Introduction}
The accurate determination of wire tensions in drift chambers is a
necessary quality control step in the construction of properly
performing chambers. Numerous techniques are described in the
literature\cite{old_tricks} that rely on a variety of means to induce
oscillations in the wire being measured. Typically, these oscillations
induce an emf or change the mutual capacitance between the anode and
its neighboring electrodes. These induced effects are then maximized
by explicitly tuning the frequency of the input perturbation.  The
frequency for which the response is maximum is then related to the
wire tension by a simple expression for the frequency of possible
standing waves on a stretched wire as a function of the wire's
tension. Such techniques, while useful, have the drawbacks that the
anode wire be electrically continuous and that the input perturbation
be explicitly tuned in frequency to maximize the response.

We have developed an automated wire tension measuring scheme that is
suitable for drift chambers with electrically discontinuous anode
wires and that requires no time-consuming frequency adjustment of the
input perturbation. Electrically discontinuous anode wires, formed by
fusing a dielectric to two halves of an anode wire, are an expedient
occupancy reducing measure for chambers exposed to a large flux of
charged particles.  Our technique was developed to measure the
tensions of the electrically discontinuous anodes of the
$\simeq56,000$ channel forward straw tracker of the BTeV project at
FNAL. Other detectors, such as the transition radiation tracker of the
ATLAS experiment at LHC, also use drift tubes with electrically
discontinuous anodes.

\section{Principle of the method}

The general idea of our technique is to use a short burst of sound
with a uniform spectral density over some finite frequency range to
excite standing waves on a biased anode wire and to then measure the
resulting power spectral density (PSD) associated with the induced
current sourced from the anode-cathode capacitive system. The PSD, as
shown below, peaks at a frequency that corresponds to the tension of
the anode. Exciting the wire with a spectrum of frequencies
simultaneously rather than serially with a single frequency allows the
tension measurement to be done accurately with great speed.

It is well known that a wire of uniform linear mass density $\mu$ under
tension $T$ with fixed end-points a distance $L$ apart supports standing
waves of frequency 

\begin{equation}
\hfill f_n={n\over2 L}\sqrt{T/\mu}\,,\label{eq:fun}\hfill
\end{equation}

\noindent where $n$ is a positive integer and labels the $(n-1)$th harmonic.
As a biased anode vibrates with respect to its neighboring cathode(s),
the resulting change in the mutual capacitance between the two
produces a current $\dot{Q}(t)= \dot{C}(t)V_0$, where $C(t)$
is the anode-cathode mutual capacitance and $V_0$ is the constant bias
voltage of the anode with respect to the cathode. The current
$\dot{Q}(t)$ can be processed by a simple transimpedance amplifier to
produce an output voltage $V^{\prime}(t)$ which in turn can be fourier
analyzed to produce its associated PSD. The frequency at which the PSD peaks is
then converted to the anode tension $T$ by Eq.~\ref{eq:fun}.

A sound burst with the desired uniform spectral composition is easily
generated using the basic properties of the fourier transform of a
function in the time domain $h(t)$ and its conjugate in the frequency
domain $H(f)$:
\begin{equation}
H(f)= \int^{\infty}_{-\infty} h(t)\e^{2\pi {\mathrm i}ft}\d t \label{eq:hf}
\end{equation}
\begin{equation}
h(t)= \int^{\infty}_{-\infty} H(f)\e^{-2\pi {\mathrm i}ft}\d f, \label{eq:ht}
\end{equation}

\noindent where the corresponding ``one-sided'' PSD of the function
$h(t)$ is defined as
\begin{equation}
P_h(f)\equiv |H(f)|^2 + |H(-f)|^2\ \ \ 0\leq f < \infty. \label{eq:psd}
\end{equation}

If we further define a constant $H(f)$, within an overall scaling factor, as
\begin{equation}
H(f)= \left\{\begin{array}{ll}
    \pi &\mathrm {if}\ a_1/(2\pi) < |f| < a_2/(2\pi)\\
    0   & \mathrm{otherwise,}
    \end{array}
\right. \label{eq:uniform_hf}
\end{equation}

\noindent where $a_1$ and $a_2$ are constants, then from
Eq.~(\ref{eq:ht}) the appropriate voltage waveform $h(t)$, again
within an overall scale factor, to feed a sound speaker to produce our
sound burst is:

\begin{eqnarray}
h(t) & = & \sin(a_2 t)/t - \sin(a_1 t)/t \\
     & = & a_2\,\mathrm {sinc} (a_2 t/\pi) - \mathrm {sinc}(a_1 t/\pi). \label{eq:burst_ht}
\end{eqnarray}

\section{Procedure and test results}

We demonstrate our tension measuring technique using a representative
prototype straw tube from the BTeV project. The straw tube is
comprised of a $20\,\mu$m diameter gold-plated tungsten anode wire
inside a $4\,$mm diameter kapton straw with a conductive inner
surface. The anode is centered inside the $100\,$cm long straw using
special helical fixtures that set the radial position of the anode at
both straw ends as well as at the straw mid-point. The distance
between neighboring fixtures, corresponding to node locations for
standing waves, is $L=50\,$cm. A simple pulley and hanging mass system
allows the anode to be tensioned with different mass values. The anode
is biased at $V_0=70\,$V with respect to the cathode, far less than
the typical bias value used for actual tracking operation. The front
face of a $1.5\,$Watt personal computer (PC) speaker used to vibrate
the anode is positioned $5\,$cm above the straw mid-point and the
volume level is set at 30\% of maximum.

The sound generation and output signal processing are done using a
combination of commercial hardware and software (LabVIEW), an op-amp
configured as a transimpedance amplifier and the speaker. A standard
LabVIEW routine calculates the sinc~function of Eq.~\ref{eq:burst_ht}
using values of $a_1$ and $a_2$ that make $H(f)$ loosely bracket the
frequency of the fundamental mode corresponding to the anode wire's
anticipated tension. For example, in the case of the anode tensioned
to $50\,$gm, $f_1\simeq 225\,$Hz so that $a_1$ and $a_2$ can be set to
correspond to $f= 200\,$Hz and $f=250\,$Hz, respectively.

\begin{figure}
\begin{center}
\includegraphics*[width=10cm]{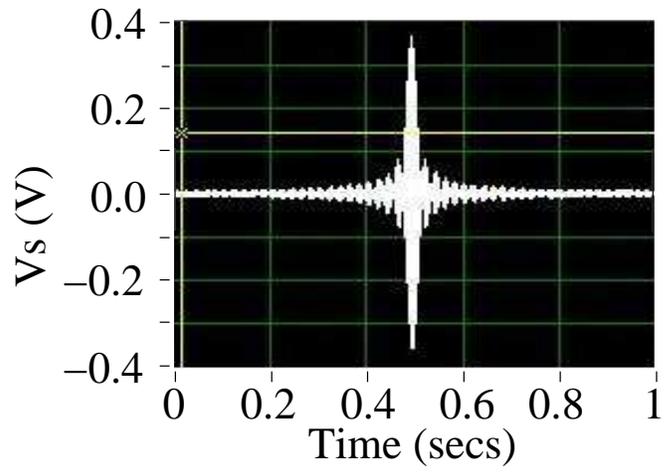}
\end{center}
\caption{The speaker voltage $V_s$ as a function of time over its
1-second duration.}
\label{fig:sinc}
\end{figure}

\begin{figure}[h]
\begin{center}
\includegraphics*[width=10cm]{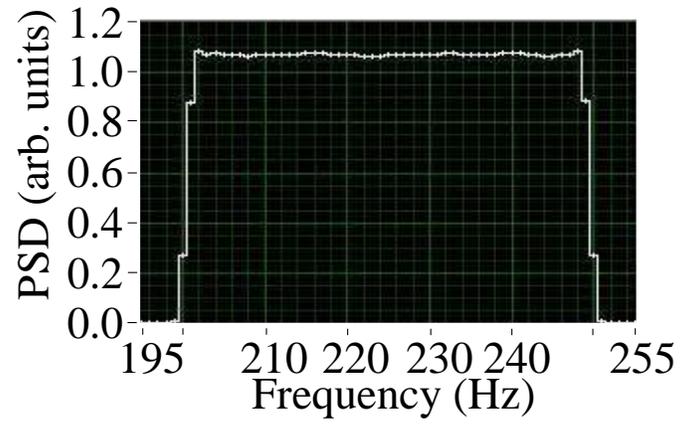}
\end{center}
\caption{The one-sided PSD of the speaker voltage. Note the flat top
and the sharp edges centered around the nominal fundamental frequency
($225\,$Hz) of the tensioned anode.}
\label{fig:spkrpsd}
\end{figure}

We generate a speaker voltage proportional to $h(t)$ by first software
sampling $h(t)$ (at a rate four times faster than what we eventually
sample our final output signal with) and then using a
digital-to-analog converter (DAC) located on a National Instruments
DAQ card\footnote[1]{NI PCI-6036E} to feed the speaker a sequence of
voltage levels. The duration of the voltage waveform is 1~second and
its quality can be measured by using an analog-digital-converter (ADC)
located on the same DAQ card.  Fig.~\ref{fig:sinc} shows the speaker
voltage and its apparent sinc-like shape. After digitization, the
corresponding one-sided PSD of the speaker voltage is computed using
the fast fourier transform (FFT) technique contained in a standard
LabVIEW routine. The result is shown in Fig.~\ref{fig:spkrpsd},
showing clearly its flat-top nature and the desired bracketing of the
anode's nominal fundamental frequency.

\begin{figure}[ht]
\begin{center}
\includegraphics*[width=13cm]{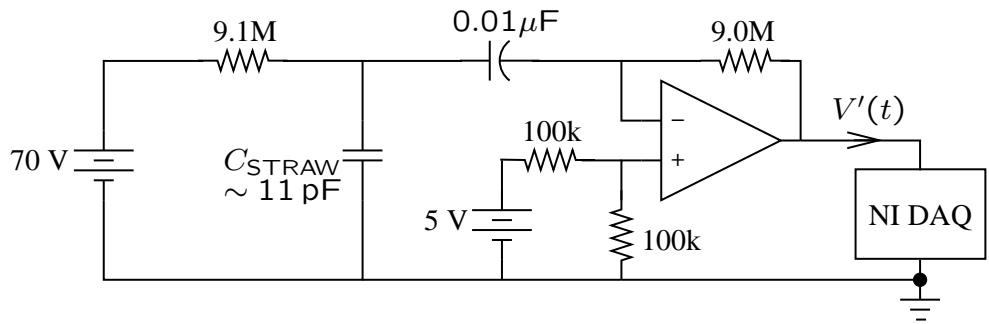}
\end{center}
\caption{Block diagram of the straw tube readout circuit. NI DAQ
refers to a commercial data acquisition card and a
special purpose connector (NI BNC-2110) from National Instruments Corp.}
\label{fig:ro}
\end{figure}

\begin{figure}[t]
\begin{center}
\includegraphics*[width=10cm]{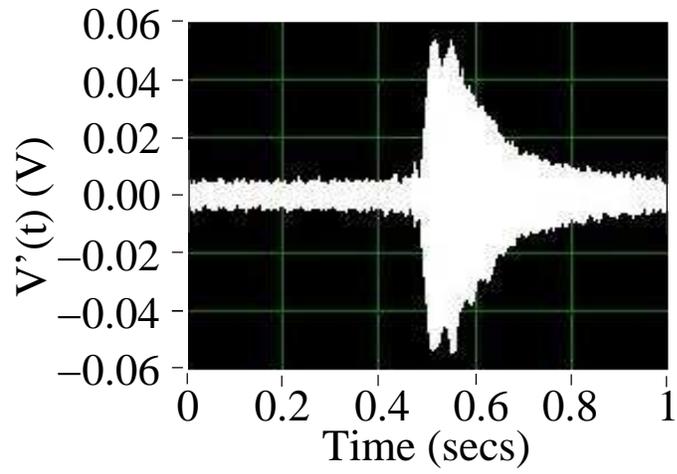}
\end{center}
\caption{Envelope of the digitized output signal $V^{\prime}(t)$ over the 1-second
duration of the sound burst.}
\label{fig:vprime}
\end{figure}

\begin{figure}[t]
\begin{center}
\includegraphics*[width=10cm]{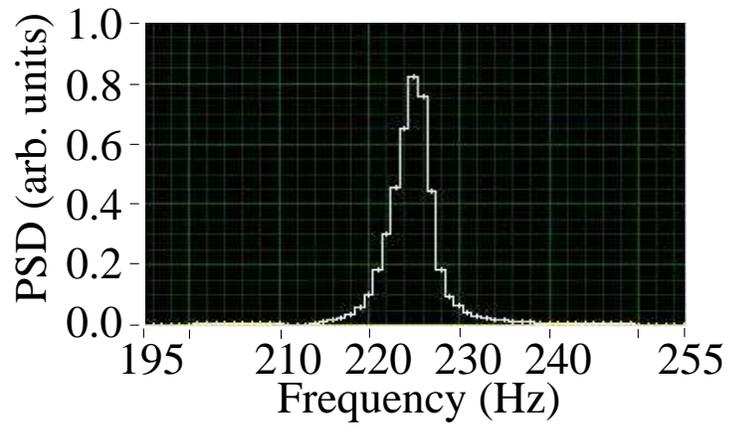}
\end{center}
\caption{Typical one-sided PSD of the output voltage $V^{\prime}(t)$
as a function of frequency. The vertical scale is arbitrary. The peak
corresponds to the fundamental frequency of the tensioned anode wire.}
\label{fig:vp_psd}
\end{figure}

The equivalent circuit of the biased straw tube and its readout
circuitry is shown in Fig.~\ref{fig:ro}. The FET-input op-amp is a
Burr-Brown OPA137P and the large feedback resistor is required because
the straw's induced current is quite small. (The estimated fractional
change in the straw's capacitance due to anode wire vibration is
$\delta C/C\sim 10^{-3}$. See Ref.~\cite{cap}.) The op-amp's output
voltage $V^{\prime}(t)$ is digitized by the DAQ card's ADC (we
uniformly sample 10,000 times during the 1-second sound burst) and
then fourier analyzed using the FFT algorithm. The computation time
for the FFT is negligible.  The digitized $V^{\prime}(t)$ is shown in
Fig.~\ref{fig:vprime} and its PSD as a function of frequency is shown
in Fig.~\ref{fig:vp_psd}.  A clear peak at the nominal fundamental
frequency of the anode is seen. The total time for sound generation
and signal processing is 2-$3\,$seconds per anode.

Many systematic tests were performed to verify the robustness of the
technique. Altering the speaker position along the straw length and
changing the horizontal orientation of the anode to vertical had no
appreciable effect on the peak frequency of the output voltage PSD.
The peak frequency was similarly insensitive to a reduction of the
anode-cathode bias voltage by a factor of two. Multiple measurements,
made by detaching and reattaching the same $50\,$gm hanging mass for
each trial, changed the PSD's peak frequency by $\sim 1.5\,$Hz.  We
estimate the total systematic error in the measured fundamental
frequency to be $\sim 2$~Hz.  The dominant systematic error in
determining the anode tension is the uncertainty in the precise value
of $L$ to use in Eq.~\ref{eq:fun}.  We estimate this to be
equivalent in our prototype to an additional frequency error of $\sim
3\,$Hz.  For comparison, a $5\,$Hz frequency error translates to a
$2\,$gm error at $50\,$gm nominal tension.

\begin{figure}[t]
\begin{center}
\includegraphics*[width=14cm]{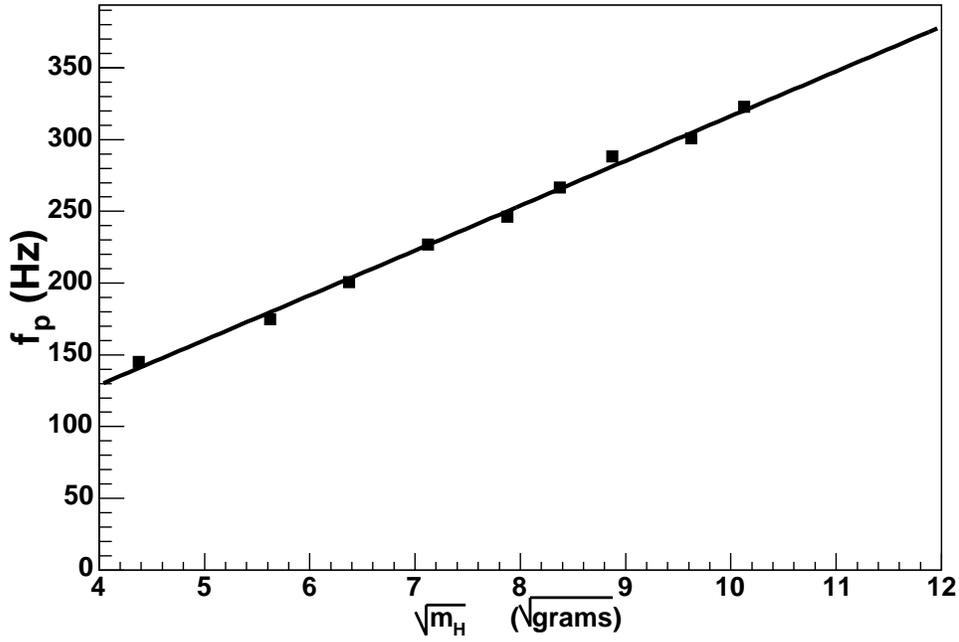}
\end{center}
\caption{Peak frequency $f_p$ of the output voltage PSD as a function
of the square root of the hanging mass $m_H$. Each data point is the
average of 20 individual measurements. (The error bars are smaller
than the data markers.) The heavy line is a linear fit of $f_p$ to
$\sqrt{m_H}$.}
\label{fig:f_v_t}
\end{figure}

The dynamic range of the technique was investigated by hanging
different masses in the range 20-$100\,$gm from the anode, producing a
nominal anode fundamental frequency in the range 145-$320\,$Hz. For
each nominal tension, the above procedure was repeated and the peak
PSD frequency determined. Each resulting PSD plot was similar in shape
to Fig.~\ref{fig:vp_psd}.  The cumulative results are shown in
Fig.~\ref{fig:f_v_t} where the measured peak PSD frequency $f_p$ (each
data point is an average of 20 individual measurements) is plotted as
a function of the square root of the hanging mass $\sqrt{m_H}$. The
heavy line is a linear fit of $f_p$ to $\sqrt{m_H}$, with a
$\chi^2/$degree-of-freedom $=1.1$, showing excellent agreement with the
form of Eq.~\ref{eq:fun}.

\begin{ack}
We thank D. Rust for useful technical discussions.
\end{ack}

\label{}

\newpage
\vspace*{10cm}
\vfill

\pagebreak
\vspace*{10cm}
\vfill
\pagebreak

\vspace*{20cm}
\vfill

\vspace*{20cm}
\vfill

\newpage

\end{document}